\documentclass[12pt]{article}
\usepackage{latexsym}
\usepackage{amssymb}
\usepackage{cite}
\usepackage{epsfig}
\title{\LARGE Leptonic Decays of the $W$-Boson in a Strong
Electromagnetic Field {\footnote{Published in: Yad. Fiz. {\bf
67}, 2116 (2004), [Phys. At. Nucl. {\bf 67}, 2095 (2004)].}}}
\author{\large A.V.Kurilin{\footnote{E-mail address:
kurilin@mail.ru}}\\
Moscow State Open Pedagogical University,\\
Verkhnyaya Radishchevskaya, ul. 16-18, Moscow, 109004 Russia\\}
\date{Received October 14, 2003}

\begin{document}
\begin{titlepage}
\maketitle \thispagestyle{empty}
\begin{abstract}{\normalsize
The probability of $W$-boson decay into a lepton and a neutrino,
$W^{\pm}\rightarrow\ell^{\pm}\tilde\nu_{\ell}$,
in a strong electromagnetic field is calculated.
On the basis of the method for deriving exact solutions
to relativistic wave equations for charged particles,
an exact analytic expression is obtained for the partial decay width
$\Gamma(\varkappa)=\Gamma(W^{\pm}\rightarrow \ell^{\pm}
\tilde\nu_{\ell})$
at an arbitrary value of the external-field-strength parameter
$\varkappa=e M_W^{-3}\sqrt{-(F_{\mu\nu}q^\nu)^2}$.
It is found that, in the region of comparatively weak fields,
($\varkappa\ll 1$) field-induced corrections to the standard decay width
of the $W$-boson in a vacuum are about a few percent.
As the external-field-strength parameter is increased,
the partial width with respect to $W$-boson decay through
the channel in question, $\Gamma(\varkappa)$, first decreases,
the absolute minimum of $\Gamma_{\rm min}=0.926\cdot\Gamma(0)$
being reached at $\varkappa=0.6116$.
A further increase in the external-field strength leads
to a monotonic growth of the decay width of the $W$-boson.
In superstrong fields ($\varkappa\gg 1$), the partial width with respect
to $W$-boson decay is greater than the corresponding partial width
$\Gamma^{(0)}(W^{\pm}\rightarrow \ell^{\pm}\tilde\nu_{\ell})$
in a vacuum by a factor of a few tens.}
\end{abstract}
\end{titlepage}

\large
\section{INTRODUCTION}
\hspace*{\parindent} Presently, the Standard Model of
electroweak interactions is the basis of our knowledge in the
realms of elementary-particle physics. Many compelling pieces of
evidence that the Standard Model describes correctly lepton and
quark-interaction processes, which occur via the exchange of
intermediate vector $W^{\pm}$ and $Z^0$ bosons, have been
obtained over the past decade. The majority of experiments
currently performed at the LEP and SLC electron-positron
colliders are devoted to studying the properties of these
particles, which mediate weak interactions \cite{CERN-96-2}.
Concurrently, the properties of $W$-bosons are being
investigated at the Tevatron proton-antiproton accelerator. The
accuracy reached in those experiments makes it possible to test
the predictions of the Standard Model at the level of radiative
corrections. For example, the upgrade of the LEP
electron-positron collider, which is now referred to as LEP-2,
enabled physicists working at CERN to observe, for the first
time, the double production of $W$-bosons, $e^+e^-\rightarrow
W^+W^-$ \cite{CERN-96-1}. This reaction provides one of the most
promising tools for precisely determining the $W$-boson mass
($M_W=80.423\pm 0.039$~GeV) and width ($\Gamma_W = 2.12\pm 0.04$
GeV) \cite{pdg-2002}. Even at present, the errors in
experimentally measuring the cross section
$\sigma(e^+e^-\rightarrow W^+W^-)$ are as small as about one
percent, whence it follows that theorists must calculate a
number of $O(\alpha)$ radiative corrections to the tree diagrams
for this process or radiative corrections of a still higher
order. It should be noted that, because of $W$-boson
instability, one actually has to deal with the more complicated
multiparticle reaction $e^+e^-\rightarrow W^+W^-\rightarrow 4f$,
where $f$ stands for the fermion products of $W$-boson decay.
Therefore, it is very difficult to perform an exact analytic
calculation of all radiative corrections to the cross section
for this process, and this will hardly be done in the near
future.

In this connection, it is reasonable to discuss alternative methods
for studying the properties of intermediate vector bosons.
In this study, we aim at calculating the effect of strong electromagnetic
fields on the leptonic mode of $W$-boson decay. The electromagnetic
interactions of these particles, which are mediators of weak interactions,
are the subject of special investigations \cite{Kurilin-1999}.

The point is that the general form of the Lagrangian describing
electromagnetic $\gamma WW$ interactions and satisfying the requirements of
$C$ and $P$ invariance has not yet been obtained experimentally.
From the theoretical point of view, it is quite admissible to extend
the Standard Model by including in it new-physics effects that
would generalize the minimal $\gamma W W$ vertex via the
introduction of two dimensionless parameters
$k_\gamma$ and $\lambda_\gamma$ \cite{Hagiwara-87},

\begin{eqnarray}
{\cal{L}}_{\gamma W
W}=-iek_{\gamma}F^{\mu\nu}W^+_{\mu}W^-_{\nu}-
ieW^+_{\mu\nu}W^{-\mu}A^{\nu}+ieW^-_{\mu\nu}W^{+\mu}A^{\nu}+
\nonumber\\
+ie(\lambda_{\gamma}/M_W^2)W^+_{\mu\alpha}W^{-\mu\beta}
F^{\alpha}_{\ \beta},
\end{eqnarray}
where
$W^{\pm}_{\mu\nu}=\partial_{\mu}W^{\pm}_{\nu}
-\partial_{\nu}W^{\pm}_{\mu}, \quad F_{\mu\nu}=\partial_{\mu}A_{\nu}
-\partial_{\nu}A_{\mu},$
and $\quad A^{\nu}$
is the 4-potential of the electromagnetic field being considered.
In the Standard Model of electroweak interactions due
to Glashow, Weinberg, and Salam, these parameters have the following
values at the tree level:
$k_\gamma=1$, $\lambda_\gamma=0$.
A precise experimental verification of these conditions must be
performed by studying the reaction
$e^+e^-\rightarrow W^+W^-$.
However, this problem has yet to be solved conclusively.

In view of the aforesaid, the approach to studying the gauge
structure of $W$-boson electromagnetic interactions on the basis
of an analysis of $W$-boson decays in an external field is of
particular interest. In the present study, we calculate the
probability of the reaction $W\rightarrow\ell\tilde\nu_\ell$
and the changes that this reaction induces in the total decay
width $\Gamma_W$, relying on the method employing exact
solutions to relativistic wave equations.
It should be noted that precise measurements
of the decay width of the $W$-bosons are of great interest both
for theorists and for experimentalists. This is because all
processes associated with the production of these particles at
electron-positron colliders are investigated by analyzing the
leptonic or hadronic products of $W$-boson decays.
In addition, the $W$-boson width is used as one of the parameters
that form a basis for calculating radiative corrections to
electroweak processes occurring at energies in the vicinity of
the W resonance; therefore, it is of paramount importance to
have precise theoretical predictions for the width
$\Gamma_W$.

One-loop radiative corrections to the decay width of the
$W$-boson have already been calculated in the literature.
In the approximation of massless fermions, they were obtained
in a number of studies
\cite{Marciano,Inoue-80,Consoli-83,Bardin-86,Jegerlehner,Jun-91}.
In that case, the main contribution comes from strong-interaction
effects (about $4\%$ with respect to the initial value
of $\Gamma_W$) \cite{Denner,Chang-82,Alvarez-88},
whereas the corrections from electroweak processes are quite modest.
It is noteworthy that new-physics phenomena (supersymmetry and so on)
also make rather small contributions
(see, for example, \cite{Rosner-94}).
This is also corroborated by the calculations of the $W$-boson
decay width within the two Higgs doublet model
\cite{Shin-95}.

At the same time, the possibility of studying the properties of the
$W$-bosons via changing external conditions under which their
production occurs has not yet been explored. In high-energy physics,
the method of channeling relativistic particles through single crystals,
in which case such particles are directed along the crystal axes
and planes formed by the regular set of crystal-lattice atoms
\cite{Cern-94-05}, has been known for a long time.
The electric fields generated by the axes and planes of single
crystals can reach formidable values (above $10^{10}$ V/m),
extending over macroscopic distances. This changes substantially
the physics of all processes in an external field in relation to
the analogous phenomena in a vacuum. Thus, single crystals prove
to be a unique testing ground where one can study reactions that
become possible in the presence of a strong external electromagnetic
field.

\section{PROBABILITY OF $W$-BOSON DECAY IN AN EXTERNAL FIELD}
\hspace*{\parindent}

In the leading order of perturbation theory, the matrix element
for the reaction of $W$-boson decay to a lepton $\ell$ and a
neutrino $\tilde\nu_\ell$ is given by
\begin{equation}
\label{SW-matrix} S_{fi}=\frac{ig}{2\sqrt{2}}\int d^4 x
\overline{\Psi}_\ell(x,p)\gamma^\mu (1+\gamma^5)\nu_\ell^c(x,p')
W_\mu(x,q).
\end{equation}

Here, an external electromagnetic field is included through a
specific choice of wave functions for the charged lepton $\ell$
and the W boson: $\Psi_\ell(x,p)$  and $W_\mu(x,q)$, respectively.
Within this approach, one goes beyond ordinary perturbation theory,
taking into account nonlinear and nonperturbative effects of an
external field in the probability of the reaction under consideration.
The explicit form of the wave functions for charged particles in an
external field can be obtained by solving the corresponding
differential equations that are determines by the Standard Model
Lagrangian (see, for example \cite{Kurilin-1999,Kurilin-1994}).
In the present study, we restrict our consideration to the case
of a so-called crossed field, whose strength tensor satisfies
the relations
\begin{equation}
\label{crossed} F_{\mu\nu} F^{\mu\nu}=F_{\mu\nu}\tilde
F^{\mu\nu}=0.
\end{equation}

Investigation of quantum processes in a crossed field is the
simplest way to analyze the transformations of elementary particles
in electromagnetic fields of arbitrary configuration.
This is because all formulas obtained in the semiclassical
approximation for the probabilities of processes that occur in a
crossed field are also applicable to describing analogous processes
in arbitrary constant electromagnetic fields.
Thus, the crossed-field model appears to be the most universal
in the region of relatively weak fields. At the same time,
the wave functions for charged particles in a crossed field have
the simplest form, and this renders relevant mathematical
calculations much less cumbersome. In our case, the $W$-boson and
the lepton ($\ell$) wave function are expressed in terms of the
external-crossed-field strength tensor
$F_{\mu\nu}=\partial_\mu A_\nu - \partial_\nu A_\mu$
as
\begin{eqnarray}
\label{lepton-wf} \Psi_\ell(x,p)=\exp\biggl[-ipx -
\frac{ie(pa)}{2(pFa)}(x^\mu F_{\mu\lambda} a^\lambda)^2
-\frac{ie^2}{6(pFa)}(x^\mu F_{\mu\lambda}
a^\lambda)^3\biggr]\times
\nonumber\\
\times\biggl\{1-\frac{e(xFa)}{4(pFa)}(F_{\mu\lambda}\gamma^\mu
\gamma^\lambda) \biggr\} \frac{u(p)} {\sqrt{2p_0 V}};
\end{eqnarray}

\begin{eqnarray}
\label{W-function}
W_{\mu}(x,q)=\exp\biggl[-iqx-\frac{ie(qa)}{2(qFa)}(x^\mu
F_{\mu\lambda} a^\lambda)^2 -\frac{ie^2}{6(qFa)}(x^\mu
F_{\mu\lambda} a^\lambda)^3 \biggr]\times \nonumber\\
\times\biggl\{g_{\mu\nu}-\frac{e(xFa)}{(qFa)}F_{\mu\nu} +
\frac{e^2(xFa)^2}{2(qFa)^2}(F_{\mu\lambda}a^\lambda)
(F_{\nu\sigma}a^\sigma)\biggr\}\frac{v^{\nu}(q)}{\sqrt{2q_0 V}}.
\end{eqnarray}

The spin component of these wave functions, which are
normalized to the three-dimensional volume $V$ of the space,
is determined by the Dirac bispinor $u(p)$ and the $W$-boson
polarization 4-vector $v^{\nu}(q)$.
Concurrently, it is assumed that the external-electromagnetic-
field potential $A_{\mu}(x)$ is taken in the gauge
\begin{equation}\label{gauge}
A_\mu (x)=-a_\mu(x^\alpha F_{\alpha\beta} a^\beta),
\end{equation}
where the unit constant $4$-vector $a_\mu$ satisfies the conditions
\begin{equation}\label{a-cond}
a_\mu a^\mu =-1, \qquad F_{\mu\nu}=(a_\mu F_{\nu\lambda}-a_\nu
F_{\mu\lambda})a^\lambda.
\end{equation}

Let us substitute the lepton and $W$-boson wave functions (\ref{lepton-wf})
and (\ref{W-function}) into expression (\ref{SW-matrix}) for
the $S$-matrix element and integrate $\mid S_{fi} \mid^2$
over the lepton and neutrino phase spaces.
After some simple but cumbersome algebra, we obtain
\begin{eqnarray}
\label{W-decay}
\Gamma(W\rightarrow\ell\tilde\nu_\ell \mid\varkappa)
=\frac{g^2 M_W} {48\pi^2} \int\limits_0^1 du \biggl\{
\biggl[1-\frac{(m_\ell^2 + m_\nu^2)} {2 M_W^2} - \frac{(m_\ell^2
- m_\nu^2)^2}{2M_W^4}\biggr] \Phi_1(z)-\nonumber\\
-2\varkappa^{2/3}\biggl(\frac{u}{1-u}\biggr)^{1/3}\biggl[1-2u+2u^2
+ \frac{(m_\ell^2+m_\nu^2)}{2M_W^2}\biggr] \Phi'(z) \biggr\}.
\end{eqnarray}

The partial decay width of the $W$-boson can be expressed in terms
of the special mathematical functions $\Phi'(z)$ and $\Phi_1(z)$
(see Appendix), which depend on the argument
\begin{equation}
\label{z} z=\frac{m_\ell^2 u + m_\nu^2 (1-u) - M_W^2 u(1-u)}
{M_W^2 [\varkappa u^2 (1-u)]^{2/3}}.
\end{equation}

In the semiclassical approximation, the external-electromagnetic-field
strength tensor appears in expression (\ref{W-decay}) for the partial
decay width only in a combination with the $W$-boson energy-momentum
$4$-vector $q^\mu$ via the dimensionless parameter
\begin{equation}
\label{kappa} \varkappa=\frac{e}{M_W^3}
\sqrt{-(F_{\mu\nu}q^\nu)^2}.
\end{equation}

The values of the parameter $\varkappa$ have a crucial effect
on the $W$-boson width in an external field. From formula
(\ref{W-decay}), one can see that, in the presence of an external
electromagnetic environment, the rate of the reaction
$W\rightarrow\ell\tilde\nu_\ell$ becomes a dynamical characteristic
that depends not only on the $W$-boson energy but also on external
conditions under which this decay occurs. Therefore, we can no
longer treat the $W$-boson width
$\Gamma(W\rightarrow\ell\tilde \nu_\ell)$
as a constant since, in an external field, it becomes a function
$\Gamma(\varkappa)$ of the parameter $\varkappa$.

Let us now consider asymptotic estimates of this function for various
values of the external-electromagnetic-field strength.
In the region of relatively weak fields that satisfy the
constraint $\varkappa\ll m_\ell/M_W$, the partial decay width
of the $W$-boson can be written as the sum of two terms;
of these, one is coincident with the decay width in a vacuum,
while the other is the correction induced by the
electromagnetic-field effect:
\begin{equation}
\label{gamma}
\Gamma(W\rightarrow\ell\tilde\nu_\ell \mid\varkappa)
=\Gamma^{(0)}(W\rightarrow\ell\tilde\nu_\ell) +
\Delta\Gamma(\varkappa).
\end{equation}

The quantity $\Gamma^{(0)}(W\rightarrow\ell\tilde\nu_\ell)$
is well known in the literature (see, for example, \cite{CERN-96-1}).
In the leading order of perturbation theory in the electroweak
coupling constant $g$, it is given by
\begin{eqnarray}
\label{freedecay} \Gamma^{(0)}(W\rightarrow\ell\tilde\nu_\ell)
=\frac{g^2 M_W} {48\pi} \sqrt{ \biggl[ 1-\biggl(
{m_\ell+m_\nu\over M_W}\biggr)^2\biggr]
\biggl[1-\biggl({m_\ell-m_\nu\over M_W}\biggr)^2 \biggr]}
\nonumber\times\\ \times\biggl[1-\frac{(m_\ell^2 + m_\nu^2)} {2
M_W^2} - \frac{(m_\ell^2 - m_\nu^2)^2}{2M_W^4}\biggr].
\end{eqnarray}

As for the other term in expression (\ref{gamma})
for the partial decay width, $\Delta\Gamma(\varkappa)$,
its value depends nontrivially on the lepton and neutrino masses.
In the case where the neutrino mass can be disregarded
($m_\nu=0$), the effect of an external field
on $\Delta\Gamma(\varkappa)$ is described by the relation
\begin{equation}\label{W-corr1}
\Delta\Gamma(\varkappa)=-\frac{g^2 M_W}{48\pi}\cdot{4\over 3}{\varkappa}^2
\biggl[1-\frac{13}{2}\biggl({m_\ell\over M_W}\biggr)^2
+16\biggl({m_\ell\over M_W}\biggr)^4- \frac{51}{4}
\biggl({m_\ell\over M_W}\biggr)^6\biggr].
\end{equation}
This expression is the second term in the asymptotic expansion of
the integral representation (\ref{W-decay}) for $\varkappa\rightarrow 0$.
Here, the entire dependence on the lepton mass $m_\ell$ is taken into
account exactly, while the terms proportional to $m_\nu$ are discarded.
The role of the effects induced by a nonzero neutrino mass is
noticeable only in the region of very weak fields,
$\varkappa \ll (m_\nu/M_W)^3$, in which case the expression for the
field-induced correction $\Delta\Gamma(\varkappa)$
develops a characteristic oscillating term that modifies expression
(\ref{W-corr1}) as follows:
\begin{equation}
\label{W-corr2} \Delta\Gamma(\varkappa)=\frac{g^2
M_W}{48\pi}\biggl[ {32\over\sqrt{6}}\varkappa\biggl({m_\nu\over
M_W}\biggr)^3 \cos \biggl(\frac{\sqrt{3}}{8\varkappa}{M_W\over
m_\nu}\biggl) - {4\over 3}\varkappa^2\biggr].
\end{equation}
From the estimates obtained above, it can be seen that weak
electromagnetic fields have virtually no effect on the decay
width of the $W$-boson - that is, the corrections to the probability
of the decay $W\rightarrow\ell\tilde\nu_\ell$ in a vacuum are
within the errors of present-day experiments.
However, this pessimistic conclusion is valid only for
$\varkappa\ll m_\ell/M_W$. In the region of rather strong
electromagnetic fields, the effects discussed here appear
to be significant.

\begin{figure}[t]
\setlength{\unitlength}{1cm}
\begin{center}
\epsfxsize=15.cm \epsffile{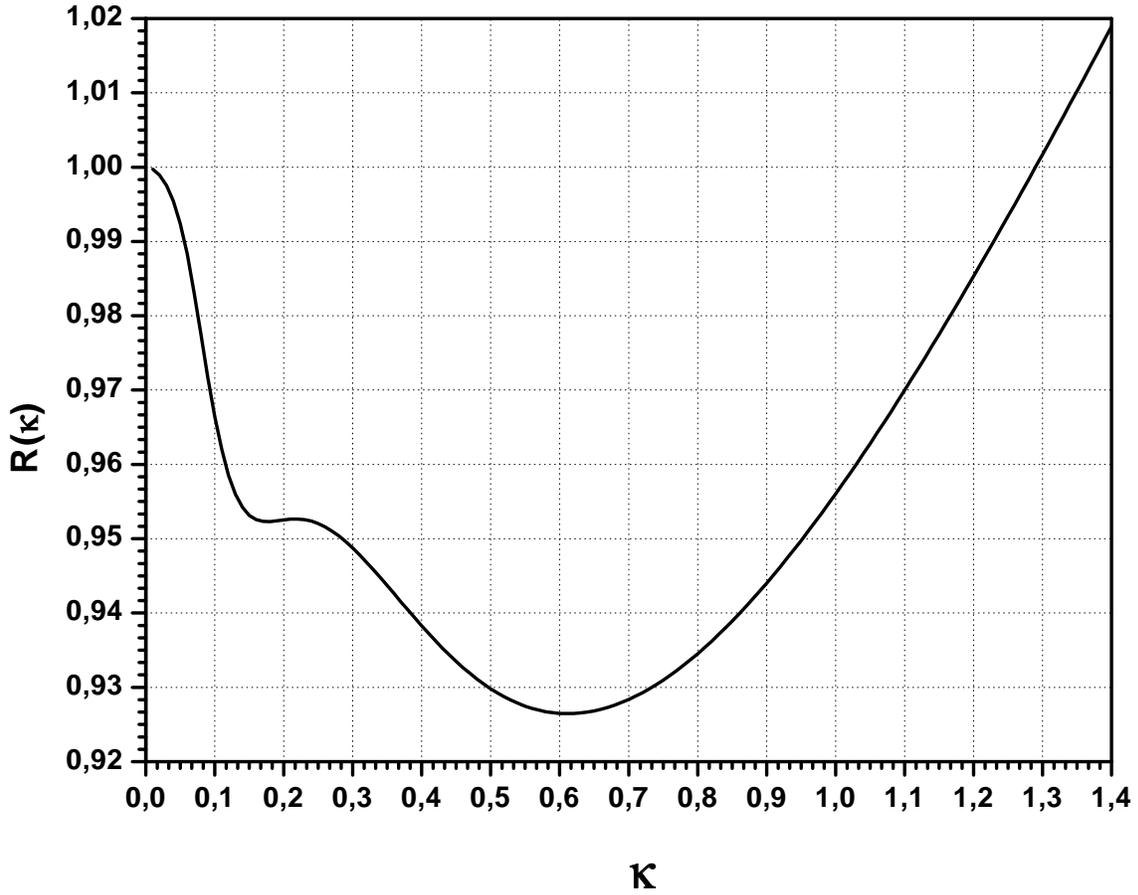}
\begin{minipage}[t]{15 cm}
\caption[]{Relative variations in the leptonic-decay width of
the $W$-boson in weak electromagnetic fields.} \label{WR1}
\end{minipage}
\end{center}
\end{figure}

\begin{figure}[t]
\setlength{\unitlength}{1cm}
\begin{center}
\epsfxsize=15.cm \epsffile{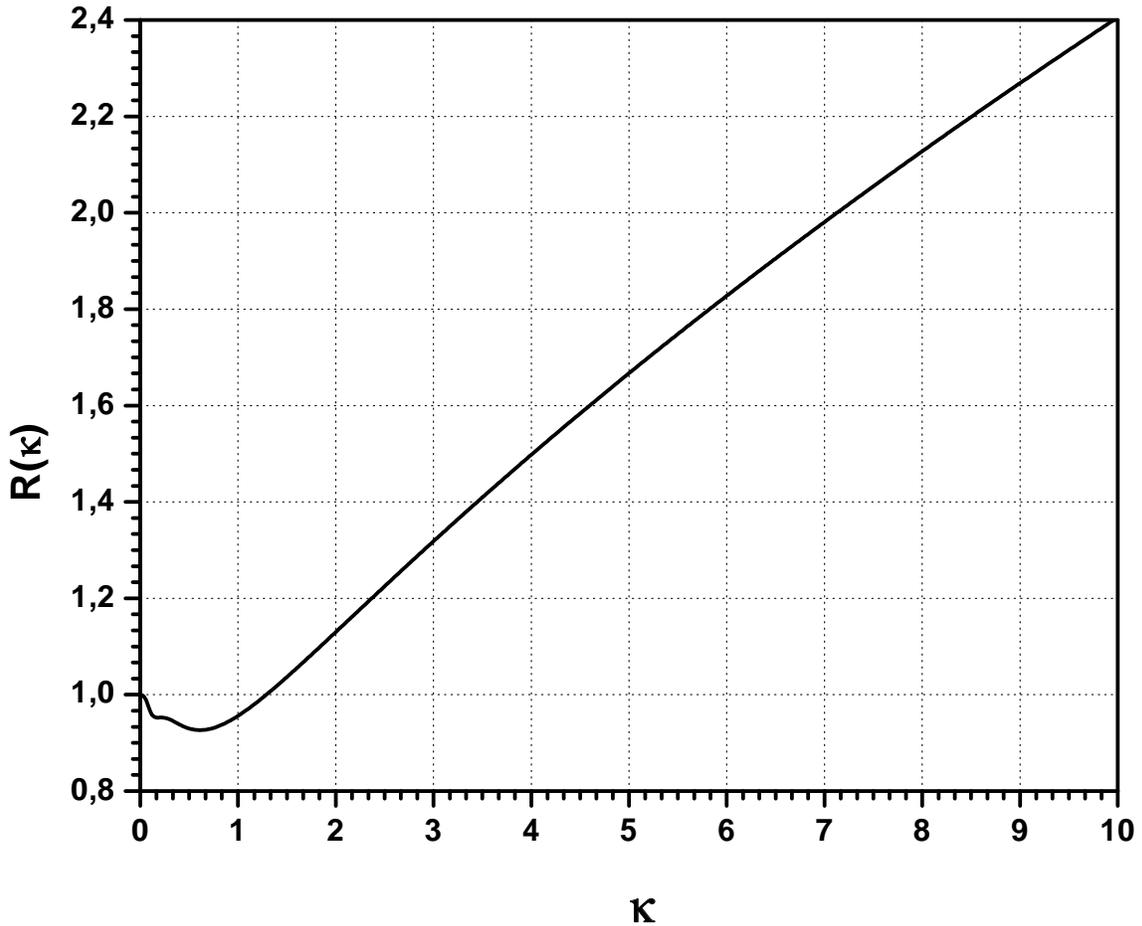}
\begin{minipage}[t]{15 cm}
\caption[] {Effect of strong electromagnetic fields on the
leptonic mode of $W$-boson decay.} \label{WR2}
\end{minipage}
\end{center}
\end{figure}

In view of this, we consider another limiting case, that of
$\varkappa\gg m_\ell/M_W$. This relationship between the
parameters makes it possible to disregard the lepton masses
against the $W$-boson mass. The relative error of this approach
does not exceed the level of corrections that are proportional
to the ratio of the masses of these particles; that is,
\begin{equation}\label{deltas}
\delta_\ell={m_\ell\over M_W}<10^{-2}; \qquad
\delta_\nu={m_\nu\over M_W}<10^{-4}.
\end{equation}
By using the approximation $\delta_\ell=\delta_\nu=0$,
one can calculate analytically the $W$-boson decay width
at any arbitrarily large value of the external-field-strength
parameter $\varkappa$. The total result of our calculations
is expressed in terms of the so-called $Gi$-function and
its derivative (see Appendix).
In order to render the ensuing exposition clearer, it is
convenient to introduce the normalized partial decay width
$R(\varkappa)$ that is defined as the ratio of the decay
width (\ref{W-decay}) in an external field to the analogous
quantity in a vacuum; that is,
\begin{equation}\label{ratio}
R(\varkappa)=\frac{\Gamma(W\rightarrow\ell\tilde\nu_\ell
\mid\varkappa)}{\Gamma^{(0)}(W\rightarrow \ell\tilde\nu_\ell)}.
\end{equation}
For the normalization factor, we chose the $W$-boson decay width
(\ref{freedecay}) calculated at zero lepton and neutrino masses:
\begin{equation}\label{freewidth}
\Gamma^{(0)}(W\rightarrow \ell\tilde\nu_\ell)\simeq \frac{G_{\rm F}
M_W^3}{6\pi\sqrt{2}}=0.227\ \mbox{ GeV}.
\end{equation}
The ultimate formula determining the effect of electromagnetic
fields on the leptonic modes of $W$-boson decay then takes the
form
\begin{equation}\label{k_width}
R(\varkappa)=\frac{4\pi}{81y} (19-2y^3)Gi'(y)-\frac{2\pi y}{81}
(11+2y^3)Gi(y)+{1\over 81}(103+4y^3),
\end{equation}
where the argument of the $Gi$-functions is related to the
parameter $\varkappa$ by the equation $y=\varkappa^{-2/3}$. This
expression is very convenient for computer calculations of the
rate of $W$-boson decay in an external field. The results of the
present numerical analysis that was based on the "Mathematica"
system are displayed in Fig.\ref{WR1}. The graph there
represents the quantity $R(\varkappa)$ (\ref{ratio}) as a
function of the electromagnetic-field strength ($\varkappa$).
One can see that, as the parameter $\varkappa$ increases, the
partial decay width of the $W$-boson gradually decreases. In the
intermediate region $\delta_\ell\ll\varkappa\ll 1$, this decay
width is well described by the asymptotic formula
\begin{equation}
\label{W-width1}
\Gamma(W\rightarrow\ell\tilde\nu_\ell)=\frac{g^2 M_W} {48\pi}
\biggl(1-\frac{8}{3}\varkappa^2 - {304\over 3}
\varkappa^4 -{4\over\sqrt{3}}\frac{\delta^3_\ell}{\varkappa}\biggr).
\end{equation}
It is noteworthy that the numerical coefficient of the first
correction to the $W$-boson decay width in a vacuum,
$({8/3})\varkappa^2$, is precisely two times as great as the
analogous coefficient of $\varkappa^2$ in the region
$\varkappa\ll\delta_\ell$ [see (\ref{W-corr1}),
(\ref{W-corr2})]. This indicates that the partial decay width of
the $W$-boson in an external field takes a minimum value in the
region around $\varkappa \sim 1$. A computer calculation shows
that the quantity $R(\varkappa)$ reaches the absolute minimum of
$R_{\rm min}=0.926$ at $\varkappa_{\rm min}=0.6116$. Thus, we
can state that, in weak fields, the deviation of the partial
decay width of the $W$-boson from that in a vacuum does not
exceed $7.4\%$. A further increase in the external-field
strength leads to a monotonic growth of the probability of
$W$-boson decay, so that, at $\varkappa > 1.3$, the decays to a
lepton and a neutrino occur faster than those in a vacuum
($R(\varkappa) > 1$). In this region of relatively strong
fields, the mean lifetime of the $W$-boson decreases sizably,
which is illustrated by the graph in Fig.\ref{WR2}. One can see
that, even at $\varkappa=7$, the dynamic al width of the
$W$-boson becomes two times larger than the static vacuum value
in (\ref{freedecay}). In superstrong fields ($\varkappa\gg 1$),
the partial decay width of the $W$-boson can be estimated with
the aid of the asymptotic expression
\begin{equation}
\label{W-width2}
\Gamma(W\rightarrow\ell\tilde\nu_\ell)=\frac{g^2 M_W} {48\pi}
\biggl\{\frac{38}{243}\Gamma(2/3)(3\varkappa)^{2/3} +
\frac{1}{3}+\frac{8}{81}\frac{\Gamma(1/3)}{(3\varkappa)^{2/3}}
\biggr\}.
\end{equation}
\section{CONCLUSION}
\hspace*{\parindent}
The effect of strong electromagnetic fields on the leptonic
mode of $W$-boson decay has been investigated. We have revealed
that, in an external field, the partial width with respect to the
decay $W\rightarrow\ell\tilde\nu_\ell$ is a nonmonotonic function
of the field-strength parameter $\varkappa$ (\ref{kappa}).
In particular, there is a domain of field-strength values where
the decays of the $W$-boson occur somewhat more slowly than
those in a vacuum. This fact has a significant effect on the
total decay width of the $W$-boson, since, in the approximation
of massless leptons and quarks, this width is known to be related
to the partial width with respect to the leptonic decays considered
here by the equation
\begin{equation}\label{Gamma-t}
\Gamma_W \simeq(N_\ell+N_q N_c)\cdot\Gamma(W\rightarrow
\ell\tilde\nu_\ell)=12\cdot\Gamma(W\rightarrow
\ell\tilde\nu_\ell),
\end{equation}
where $N_\ell=N_q=3$ are the numbers of the lepton and quark
generations and $N_c=3$ is the total number of color quarks.
Thus, one can state that, in relatively weak electromagnetic fields,
the maximum deviation of the total width of the $W$-boson from its
vacuum value $\Gamma_W$ is $7\%$. As for the region of strong
fields ($\varkappa \ge 1$), a stable trend toward an increase
in the rate of $W$-boson decays is observed here, which reduces
their lifetime to a still greater extent.
It is noteworthy that a similar nonmonotonic dependence on
the external-field-strength parameter is observed in the decay
of the scalar pion to a lepton pair,
$\pi \rightarrow\ell\tilde\nu_\ell$ \cite{Ritus-69}
This circumstance is explained by the similarity of kinematical
conditions under the which the decays of massive charged particles
proceed in a crossed field and by the fact that the reactions
$W\rightarrow\ell\tilde\nu_\ell$ and
$\pi \rightarrow\ell\tilde\nu_\ell$ are both energetically allowed;
therefore, they are possible even in the absence of external fields.
At the same time, it should be noted that the electromagnetic
interactions of the $W$-bosons are much more complicated than the
interactions of scalar pions, this leading to a number of interesting
phenomena. It is well known (see, for example, \cite{Nielsen})
that the energy spectrum of $W$-bosons in a superstrong electromagnetic
field involves the so-called tachyon mode, which is due to their
anomalous magnetic moment $\Delta\mu_W=ek_\gamma /2M_W$.
This results in that the $W$-boson vacuum becomes unstable within
perturbation theory, so that there arises, in super-strong fields,
the possibility of a phase transition to a new ground state,
this phase transition being accompanied by the restoration of
$SU(2)$-symmetry, which is spontaneously broken under ordinary
conditions \cite{Salam-75,Savvidy,Skalozub,Ambjorn}. The trend
toward the rearrangement of the $W$-boson vacuum manifests itself
in a singular behavior of many physical quantities in the vicinity
of the critical external-field-strength value of
$F_{\rm cr}=M_W^2/e \simeq 1.093\times 10^{24}$ G. For example,
the anomalous magnetic moment of Dirac neutrinos that is due to
the virtual production of $W$-bosons has a logarithmic singularity
in a superstrong field for
$F=\sqrt{F_{\mu\nu} F^{\mu\nu}/2}\rightarrow F_{\rm cr}$
\cite{Borisov-85}. Unfortunately, the crossed-field model, which
was used in the present study, is inadequate to the problem of
calculating the behavior of the $W$-boson decay width at external-field
strengths close to the critical value $F_{\rm cr}$. In order to solve
this problem, it is necessary to go beyond the semiclassical approximation
(\ref{crossed}) and to employ more complicated wave functions for
charged particles. At any rate, it seems that this problem is of
particular interest and that it deserves a dedicated investigation.

Yet another interesting result obtained in the present study is that
which concerns the effect of a nonzero mass of Dirac neutrinos on
the $W$-boson decay width in weak electromagnetic fields.
In the case of $m_\nu\ne 0$, it has been found that the correction
$\Delta\Gamma(\varkappa)$ to the vacuum decay width develops a
nontrivial oscillating term (\ref{W-corr2}), which can serve as
some kind of indication that massive neutrinos exist.
It should be noted that oscillations of the probabilities of quantum
processes in an external field emerge for a number of reactions
allowed in a vacuum if the participant particles have a nonzero
rest mass. For example, similar oscillations arise in the cross
section for the process $\gamma e\rightarrow W\nu_e$ if the
respective calculations take into account effects of a nonzero
neutrino mass \cite{Kurilin-1988}.

\newpage
\begin{center}
\section*{APPENDIX}
\end{center}
\hspace*{\parindent}

In the present study, we have employed special mathematical
functions generically termed Airy functions. A somewhat different
notation is used for Airy functions in the physical and in the
mathematical literature. Mathematicians describe Airy functions
with the aid of the symbols $Ai(z), Bi(z)$, and $Gi(z)$, which are
related to our notation as follows:
$$
\Phi(z)=\pi Ai(z), \qquad
\Phi'(z)=\pi Ai'(z), \eqno \mbox{(A.1)}
$$
$$
\Phi_1(z)=\pi^2\Bigl[Ai(z) Gi'(z)-Ai'(z) Gi(z)\Bigr].
\eqno \mbox{(A.2)}
$$
Airy functions are particular solutions to the second-order linear
differential equations
$$
Ai''(z)- z Ai(z)=0, \qquad Gi''(z)-zGi(z)=-{1\over\pi},
\eqno \mbox{(A.3)}
$$
These solutions can be represented in the form of improper integrals
of trigonometric functions,
$$
Ai(z)={1\over\pi}\int\limits_0^\infty dt\cos(zt+t^3/3),
\eqno \mbox{(A.4)}
$$
$$
Gi(z)={1\over\pi}\int\limits_0^\infty dt\sin(zt+t^3/3).
\eqno \mbox{(A.5)}
$$
The properties of the Airy functions are well known, and a compendium
of these properties can be found in mathematical handbooks
(see, for example, \cite{Abramovitz}).

\end{document}